\documentclass[aps,prb,twocolumn,amsmath,amssymb]{revtex4}             % DO NOT DELETE THIS LINE
\RequirePackage{graphicx}

\newcommand{ \etal }{\mbox{\sl et al. }}

\newcommand{ \degree }{\mbox{$^{\circ}$}}

\newcommand{\micron}{\mbox{$\mu$m}}
\newcommand{\first}{\mbox{1$^{st}$}}
\newcommand{\third}{\mbox{3$^{rd}$}}

\newcommand{\ltwo}{\mbox{L$_2$}}
\newcommand{\lthree}{\mbox{L$_3$}}

\newcommand{\mfour}{\mbox{M$_4$}}
\newcommand{\mfive}{\mbox{M$_5$}}

\begin{document}                  % DO NOT DELETE THIS LINE

     %-------------------------------------------------------------------------
     % The introductory (header) part of the paper
     %-------------------------------------------------------------------------

     % The title of the paper. Use \shorttitle to indicate an abbreviated title
     % for use in running heads (you will need to uncomment it).

\title{Resonant scattering and diffraction beamline P09 at PETRA III}
%\shorttitle{Beamline P09}

     % Authors' names and addresses. Use \cauthor for the main (contact) author.
     % Use \author for all other authors. Use \aff for authors' affiliations.
     % Use lower-case letters in square brackets to link authors to their
     % affiliations; if there is only one affiliation address, remove the [a].

%\cauthor{J.}{Strempfer}{joerg.strempfer@desy.de}
\author{J. Strempfer}
\author{S. Francoual}
\author{D. Reuther}
\author{D. K. Shukla}
\author{A. Skaugen}
\author{H. Schulte-Schrepping}
\author{T. Kracht}
\author{H. Franz}

\affiliation{Deutsches Elektronen-Synchrotron (DESY), 22603 Hamburg, Germany}
\date{\today}
%\shortauthor{J. Strempfer et al.}

     % Use \vita if required to give biographical details (for authors of
     % invited review papers only). Uncomment it.

%\vita{Author's biography}

     % Keywords (required for Journal of Synchrotron Radiation only)
     % Use the \keyword macro for each word or phrase, e.g. 
     % \keyword{X-ray diffraction}\keyword{muscle}

     % PDB and NDB reference codes for structures referenced in the article and
     % deposited with the Protein Data Bank and Nucleic Acids Database (Acta
     % Crystallographica Section D). Repeat for each separate structure e.g
     % \PDBref[dethiobiotin synthetase]{1byi} \NDBref[d(G$_4$CGC$_4$)]{ad0002}

%\PDBref[optional name]{refcode}
%\NDBref[optional name]{refcode}

\begin{abstract}
The resonant scattering and diffraction beamline P09 at PETRA III is
designed for X-ray experiments requiring small beams, energy
tunability, variable polarization and high photon flux. It is highly
flexible in terms of beam size and offers full higher harmonic suppression. A
state of the art double phase retarder setup provides variable linear
or circular polarization. A high precision Psi-diffractometer and a
heavy load diffractometer in horizontal Psi-geometry allow the
accommodation of a wide variety of sample environments. A 14 T
cryo-magnet is available for scattering experiments in magnetic
fields.
\end{abstract}

\maketitle                        % DO NOT DELETE THIS LINE

     %-------------------------------------------------------------------------
     % The main body of the paper
     %-------------------------------------------------------------------------
     % Now enter the text of the document in multiple \section's, \subsection's
     % and \subsubsection's as required.

\section{Introduction}

P09 is optimized for resonant and non-resonant elastic
scattering and diffraction experiments in the tender to hard X-ray
range. Two diffractometers are located in two different experimental hutches:
one for standard sample environments such as small cryostats or
furnaces and another one for heavy sample environment setups. A 14~T
DC magnet is also available. It is equipped with a variable
temperature insert (VTI) and an optional $^3$He probe.
In a third experimental hutch, beamline P09 hosts the hard X-ray
photo-electron spectroscopy (HAXPES) station. This paper will focus exclusively on
the resonant scattering and diffraction (RSD) stations.

The beamline is dedicated to resonant X-ray diffraction (RXD) experiments,
in which the scattering cross section is strongly polarization dependent.
However, various scattering and diffraction experiments in
need of high photon flux, variable energy, variable polarization,
focused beam are also possible.  This includes investigation of reflectivities or
superlattice reflections as a function of temperature, electric or
magnetic fields in bulk samples, thin films or multilayer systems
using linear or circular polarization, as well as in-situ
characterization of sample growth in heavy chambers. With the highly
focused X-ray beam, diffraction experiments from small single crystals
or single domains within a larger multi-domain sample are possible.

The RXD technique is nowadays widely used in various
fields \cite{Vet12} and has been applied extensively to the
investigation of electronic order in solids \cite{Bea12}. By tuning
the X-ray energy close to an absorption edge, virtual multipole
transitions are induced and the scattering process becomes sensitive
to the intermediate states into which the core electrons are
excited. From this a strong sensitivity to local environments and
asymmetries in the electron distribution arises which can have various
origins, such as charge, spin or multipolar order and can give rise
to diffracted intensity at structurally forbidden positions in
reciprocal space or an enhancement of weak magnetic reflections.

The anisotropic tensor of susceptibility (ATS) \cite{Dmi83} describes
the polarization dependent nature of the diffracted intensities at
forbidden reflections at the absorption edges. ATS scattering
originally was applied exclusively to the investigation of purely
structural anisotropies.  Independent from that, magnetic RXD was
discovered \cite{Gib88b}, which yields, dependent on the type of
transitions, strong resonant enhancement of the weak magnetic
intensities of magnetic superlattice reflections. The sensitivity of
RXD to magnetic order in the following years was extensively used to
investigate magnetic ordering phenomena as a function of temperature,
external magnetic field or pressure as a complementary probe to
neutron diffraction. The element sensitivity of XRD was used to distinguish
between contributions of different elements to the magnetic ordering.
ATS scattering was later applied to orbital and charge ordering
\cite{Mur98a} and since then also to the investigation of higher
multipolar order terms.

From the original description of the resonant magnetic cross section
\cite{Blu85} until now, theory has advanced considerably and the
description of the different resonant diffraction phenomena has been
greatly unified by the so-called multipole expansion of the electric
and magnetic state of matter \cite{DiM12}. This provides a wide field
of ordering phenomena to be investigated.

At P09, all requirements for state of the art resonant and
non-resonant scattering experiments are met, which include
energy tunability, the possibility of vertical as well as
horizontal scattering, manipulation of incident polarization, analysis
of the polarization of the diffracted beam, variable beam focus and
higher harmonic suppression in the whole energy range. The beamline
covers the absorption edges in the energy range from 2.7 to 24 keV,
which include the 4d L-edges (Mo, Ru, Rh, ...), the 5f M-edges (U),
the 4f L-edges, the 3d K-edges and the 5d L-edges.  For experiments at
low photon energies, P09 is virtually windowless except for a
20~\micron\ diamond window which separates the ring from the
beamline vacuum.
%%%%%%%%%%%%%%%%%%%%%%%%%%%%%%%%%%%%%%%%%%%%%%%%%%%%%%%%%%%%%%%%%%%%

%%%%%%%%%%%%%%%%%%%%%%%%%%%%%%%%%%%%%%%%%%%%%%%%%%%%%%%%%%%%%%%%%%%%%%%%%%%%%%%

\section{Storage ring, Undulator and Front-end}

Beamline P09 is located at the 3$^{rd}$ generation synchrotron
source PETRA III at DESY, which became operational in 2009
\cite{Bal04}. The PETRA III storage ring has a circumference of 2.3
km. It accommodates 9 straight sections corresponding to 9
sectors in the experimental hall covering 1/8th of the ring
circumference. In total, 14 beamlines are installed using canted 2~m
undulators with a canting angle of 5~mrad in 5 sectors, 5~m undulators
in 3 sectors and a $2\times 5$~m undulator in the first sector.

PETRA III sets itself
apart from other synchrotron sources due to its low horizontal
emittance of 1 nm\,rad. The lower emittance directly translates into
a reduced horizontal photon source size and beam divergence and into a high
brilliance for all beamlines. PETRA III operates at 6.084~GeV ring
energy. The ring current is injected in top-up mode with 480
(eventually 960) bunch filling pattern and 100~mA ring current in
normal operation. Reduced bunch modes with 60 and 40 bunches are used
for timing experiments. 
\begin{figure}
%\scalebox{0.58}{\includegraphics{paper_fig_new3.eps}}
\scalebox{0.58}{\includegraphics{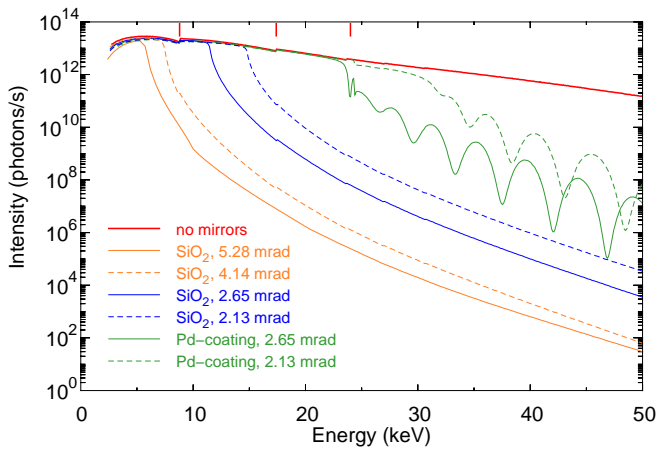}}
\caption{Intensity of direct and mirror reflected (focused) beam as a
  function of energy calculated for the U32 spectroscopy undulator,
  simulated using XTRACE \cite{See08}. The solid lines show the curves
  for a focus in the first experimental hutch (2.65 mrad and 5.28 mrad
  glancing angles) and the dashed lines the ones for a focus in the
  second experimental hutch (2.13 mrad and 4.14 mrad). The
  corresponding cut offs are given in
  Table~\ref{mirror_specs}. Vertical lines on top of the spectra mark
  transitions between undulator harmonics.}
\label{fig_mirror}
\end{figure}

P09 is located at sector 6 of the PETRA III
experimental hall. It shares the sector with beamline P08, the High
Resolution Diffraction Beamline \cite{See12}. The two beamlines are
using two separate undulators canted by an angle of 5 mrad. The
undulators are placed in a high-$\beta$ section of the storage ring,
where the calculated rms photon source sizes and divergences at 8 keV are
$140.8\times 5.2$~\micron$^2$ (H$\times$V) and $9.4\times 6.5$~$\mu$
rad$^2$ (H$\times$V), respectively. At P09, a 2 m long U32
spectroscopy undulator is used with a period length of
$\lambda_U$=31.4\,mm and a peak field of B$_0$=0.91\,T \cite{Bar08}. It generates a
continuous energy spectrum via gap scanning with a transition from \first\ to
\third\ harmonic at an energy of 8.8 keV as shown on Fig.~\ref{fig_mirror}. The
lowest gap size is 9.8 mm allowing for a minimum energy of 2.7
keV at the first harmonic. Water-cooled power slits in the beamline front-end are used to
shape the beam and to reduce the heat-load by cutting off the low
energy halo around the central cone.

\begin{figure}
\caption{Beam path through the optics of P09 (side view). The scale on
  the bottom denotes the distance in meters from the source. Left, the
  high heat-load double crystal monochromator is shown together with
  the beam stop for the white beam (BS). It is followed by the
  optional high resolution monochromator foreseen for the HAXPES
  station. Phase-retarder and focusing mirrors are also located in the
  optics hutch. On the right, optical components within the first
  (P09-EH1) and the second experimental hutches (P09-EH2) are shown up
  to the diffractometers. In the third experimental hutch, the
  position of the HAXPES spectrometer is shown.  }\label{fig_sect6}
%\scalebox{0.45}{\includegraphics{p09_optics_n.ps}}
\scalebox{0.45}{\includegraphics{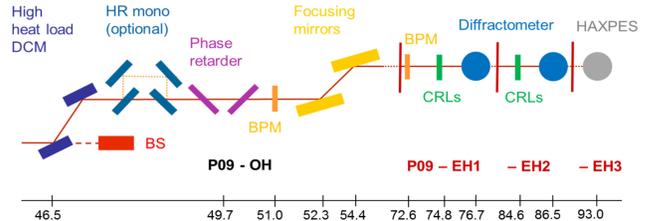}}
\end{figure}

\section{Beamline optics}

In Fig.~\ref{fig_sect6} we show the different optical elements at P09
and their distances from the source. The optics hutch P09-OH starts at
45.2~m. The first component is the high heat-load double-crystal
fixed-exit monochromator (DCM) (FMB Oxford, Osney Mead, UK) located at
46.5~m. The vertical offset of the monochromatic beam relative to the
white beam is 21~mm.  A horizontal translation of the whole
monochromator tank perpendicular to the beam allows switching between
Si(111) or Si(311) crystal pairs. Both pairs are cryogenically cooled
with liquid N$_2$. Because of the smaller Darwin width of the Si(311)
Bragg reflection, the energy resolution can be improved by almost a
factor 5 when using Si(311) instead of Si(111) at the expense of a
factor 5 in photon flux. The energy bandwidth is $(\Delta
E/E)_{(111)}=1.3\cdot10^{-4}$ and $(\Delta
E/E)_{(311)}=0.28\cdot10^{-4}$ or in absolute units 1 (0.5)~eV and 0.2
(0.1)~eV, respectively, at 8 (4)~keV.  The lowest energy at which
Si(311) crystal pairs can be used is 5.2~keV, corresponding to a
maximum monochromator angle of currently 46.7\degree. Quad beam
position monitors (QBPMs) (FMB-Oxford) at 51~m in P09-OH and at 72.6~m
in the \first experimental hutch (P09-EH1) equipped with
0.5\micron\ Ni and Ti foils can be used to stabilize the beam position
as well as its direction by adjusting position and pitch of the second
monochromator crystal.  Downstream of the monochromator at 49.7~m, the
double phase-retarder setup can be used to vary the polarization of
the incident monochromatic beam.  It is followed by two 1~m long
vertically reflecting mirrors at 52.3 and 54.4~m used for focusing and
higher harmonics rejection.  The position of the sample at the center
of rotation of the diffractometers is at 76.7~m in P09-EH1 and at
86.5~m in P09-EH2. The vacuum at the optical components is maintained
at pressures below $10^{-7}$~mbar up to the 20~\micron\ thick diamond
X-ray window (Diamond Materials, Freiburg) at the entrance to
P09-EH1. In P09-EH1, 3~m in front of the sample position, a 3$^{rd}$
300~mm long quartz mirror is available to further suppress higher
harmonics in the low energy regime below 4~keV.  Additional focusing
at fixed energy can be achieved using compound refractive lenses
(CRLs) \cite{Len99}.  In the following, the double phase-retarder
setup, the double mirror setup and the CRL setup are explained in more
detail.

\begin{figure}
\caption{ Schematic drawing of the in-vacuum double phase retarder
  setup in the optics hutch. Two individual diffraction stages with
  opposite theta and two-theta circles mounted on the roll ($\chi$)
  stages can be seen. Three diamond phase plates are mounted on each
  of the crystal changers. The inclination of the diffraction
  plane shown is $\pm 45$\degree\ for the two circles. With both
  phase plates in QWP condition this results in circular polarization
  after the first and 90\degree\ rotated linear polarization after
  the second phase plate.}\label{fig_pr}
%\scalebox{0.5}{\includegraphics{phase_retarder.ps}}
\scalebox{0.075}{\includegraphics[clip=true]{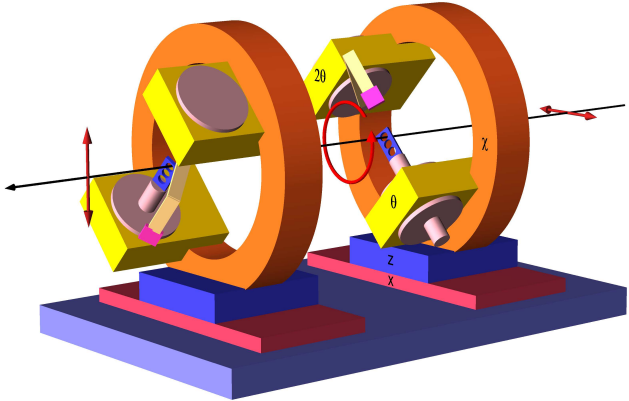}}
\end{figure}

\subsection{Phase retarder}

\begin{figure}
%\scalebox{0.075}{\includegraphics{scattconf.eps}}
\scalebox{0.075}{\includegraphics{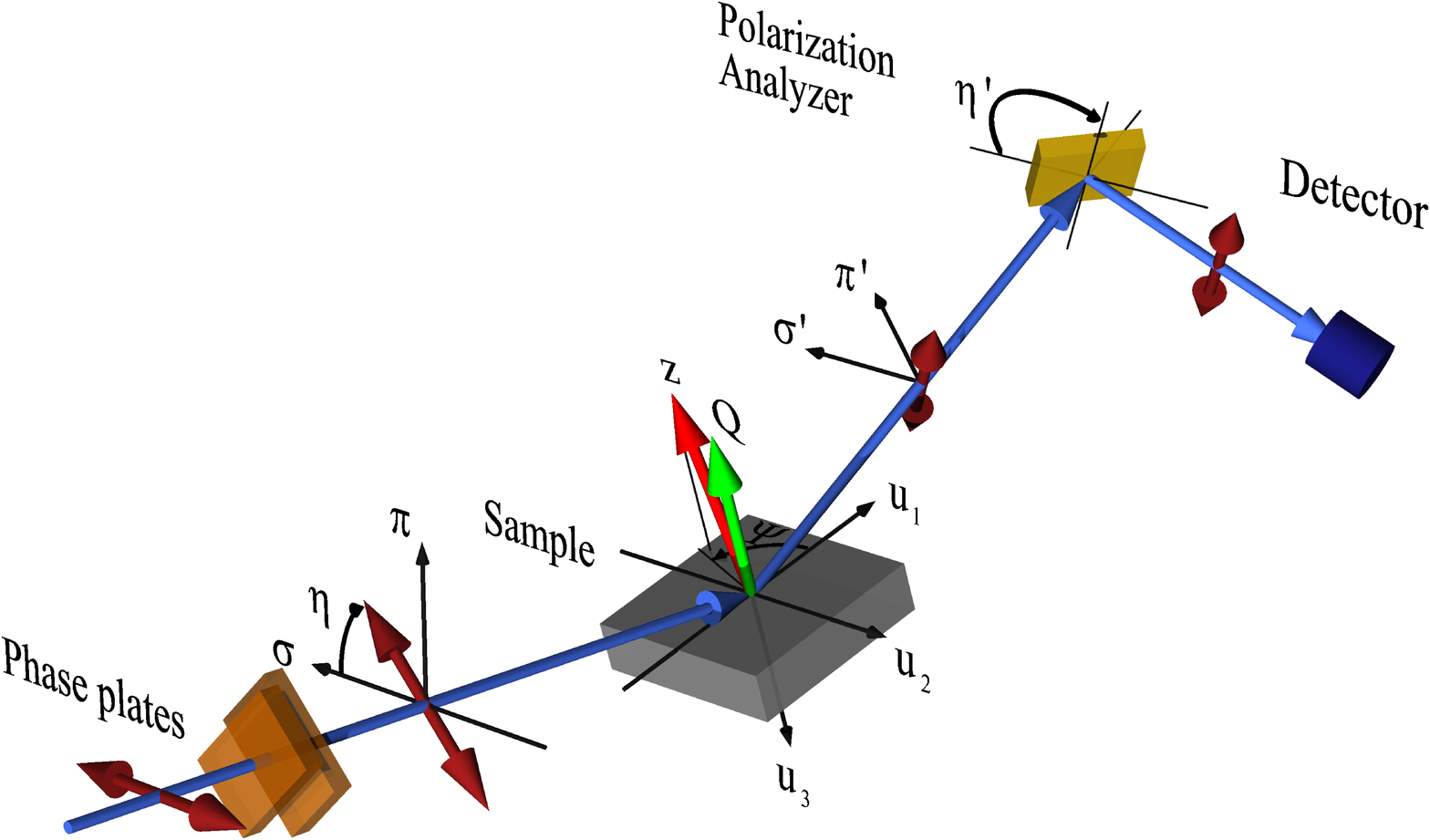}}
\caption{ Scattering configuration together with phase plates and polarization
  analyzer. The phase plates allow a rotation ($\eta$) of
  the linear incident polarization in the plane normal to the incident
  wave vector as well as generation of left and right circular
  polarization. For the sample, the scattering vector {\bf Q} and the
  uniaxial vector {\bf z} is shown, which rotates around {\bf Q} with
  the azimuth angle $\Psi$ in the plane normal to the scattering
  vector. At the sample, the reference vectors ${\bf u}_1$, ${\bf
    u}_2$ and ${\bf u}_3$ are shown \cite{Blu88}. The scattered signal
  is analyzed by the polarization analyzer, which can be rotated by
  the angle $\eta'$ in the plane normal to the scattered beam around
  the scattered wave vector. By rocking the analyzer at several
  different $\eta'$, the polarization properties of the diffracted
  beam can be determined. Polarization components perpendicular,
  $\sigma$($\sigma$'), and parallel, $\pi$($\pi$'), to the diffraction
  plane are shown for the incident (diffracted) X-ray beam.}
\label{fig_scattgeo}
\end{figure}

Magnetic structures are usually determined through the investigation
of the scattered intensity as a function of the azimuth. However such
measurements imply a rotation of the sample around the scattering
vector together with the whole sample environment ({\sl i.e.} cryostat
or magnet) over a large angular range of the azimuth at different
scattering vectors which is not always possible. Moreover, different
grains of the sample often scatter at different azimuthal angles and
thus make such measurements extremely difficult.  These difficulties
can be overcome using polarization scans 
{\sl i.e.} the
generation of incident linearly polarized X rays at a variable angle $\eta$ 
of the linear polarization plane around the beam direction and the analysis
of the dependence on $\eta$ of the Stokes parameters $P’_1$ and $P’_2$
of the diffracted signal from a sample \cite{Blu88,Det12}.
After its first implementation at beamline ID20 at ESRF \cite{Pao07}, this
method has become a routine requirement due to its
versatility. Polarization scans allow disentangling between
different contributions to the resonant scattering cross-section as
was first shown by Mazzoli {\sl et al.}\cite{Maz07}. For magnetic resonant and non-resonant scattering
experiments, full polarization scans
are used to determine the magnetic moment orientation \cite{Joh08}.
In addition, circular polarization with the
possibility of fast switching between left and right circular
polarization is of great importance, since it can be used to determine
properties of chiral magnetic structures \cite{Fab09}.  A
sketch of a diffraction experiment with phase-plates and polarization
analyzer is shown in Fig.~\ref{fig_scattgeo}. 

%%%%%%%%%%%%

Circular and rotated linear polarization can be obtained using quarter
wave plates (QWP) and half wave plates (HWP), respectively. In the
hard x-ray regime, QWP and HWP conditions can be obtained by using
perfect crystals and slightly detuning them from
the ideal Bragg condition \cite{Gil94}.  The actual phase shift
depends on the deviation angle from the Bragg position.

The double phase-retarder setup is mounted in a UHV vacuum tank and
consists of two Eulerian cradles, $\chi_{pr1}$ and $\chi_{pr2}$
separated by 340 mm (Huber Diffraktionstechnik, Rimsting, Germany)
(Fig.~\ref{fig_pr}). On each of them two high resolution rotation
stages are mounted opposite to each other acting as $\theta$ and
$2\theta$ circles. The $\theta$-circles are equipped with a $\pm$10 mm
translation making it possible to switch between 3 phase-retarding
plates.  A piezo-actuator below the plates allows oscillation with
frequencies of up to 40 Hz for fast switching the helicity of circular
polarization. On the $2\theta$ stages, passivated implanted planar
silicon (PIPS) diodes (Canberra) are located which monitor the
diffracted intensity. The two Eulerian cradles are mounted on separate
$z$- and $x$-stages. Their axes of rotation are aligned towards each
other within a precision of better than 0.1\degree. The rotation axis
of one of them can be aligned precisely along the beam axis using yaw
and pitch rotations. As a result, Bragg peaks remain in diffraction
condition within $\pm 5$ arcsec when $\chi_{pr2}$ is rotated between 0
and $-90$\degree\ and within $\pm$10 arcsec when $\chi_{pr1}$ is
rotated between 0 and $+90$\degree. 
\begin{figure}
\caption{Measured (circles) and calculated (lines) deviation angles
  for the QWP condition for a 400~\micron\ phase-plate for two
  different incident angles as a function of energy. The insert shows
  the measured variation of the Stokes parameters with varying angle
  of linear polarization $\eta$ at 6.5~keV using 2 QWPs in the
  90\degree\ geometry. $P_{lin}$ gives the degree of linear
  polarization (inv. filled triangles).}\label{fig_400mu}
%\scalebox{0.6}{\includegraphics{phase_plates_400.eps}}
\scalebox{0.6}{\includegraphics{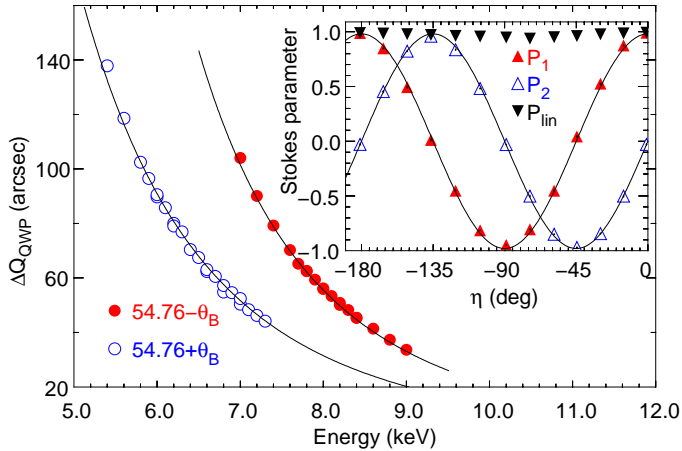}}
\end{figure}

Currently, two identical sets of synthetic type Ib diamond phase
plates are mounted on the two $\chi_{pr1}$ and $\chi_{pr2}$ stages:
two diamond plates of 100 and 200 $\mu$m thickness with the [111]
direction as surface normal and one diamond plate of 400 $\mu$m
thickness with the [100] direction as surface normal. Each plate is
oriented such that the (111) and (\=1\=11) reflections can be set in
diffraction condition, which allows an extension of the usable energy
range of each plate owing to the different glancing angles onto the
(111) planes. In Fig.~\ref{fig_400mu} the deviation angles from the
Bragg position necessary to achieve the quarter wave plate condition
(QWP) are shown for the 400~$\mu$m (100) plate for two different
glancing angles. Depending on which plate is used and which reflection
is set diffracting, the plates are either used in the asymmetric Laue
geometry or the symmetric Bragg geometry. The present set of diamond phase
plates allows manipulation of the polarization of the incident X rays
in the energy range between 3.5 and 8 keV.

Linear polarization scans are carried out using either one single HWP
or two QWPs in series and rotating their scattering planes around the
X-ray beam. Scagnoli \etal\ have shown \cite{Sca09} that using two
QWPs in series in a 90\degree\ geometry, {\sl i.e.}
$\chi_{pr1}-\chi_{pr2}= 90\degree$, allows compensation for
depolarisation effects due mostly to the energy spread, while two QWPs
in series in an antiparallel geometry, {\sl i.e.}
$\chi_{pr1}-\chi_{pr2}= 180\degree$, compensate for depolarization
effects due to the angular divergence. The angular divergence is very
low at P09 and can be neglected.  The energy spread on the other hand
contributes an angular spread of more than 10 arcsec at energies below
9~keV, which is non negligible. The combination of two QWPs in the
90\degree\ geometry results in a 5 to 10\% higher degree of linear
polarization $P_{lin}$ than (i) using one single HWP of half the total
effective thickness \cite{Fra13} and (ii) using two QWPs in the
parallel geometry, {\sl i.e.}  $\chi_{pr1}=\chi_{pr2}$
\cite{Sca09,Oki01,Ina13}.  An example of the direct beam polarization
properties for two QWPs in series in the 90\degree\ geometry at the Mn
K-edge energy is given in the inset of Fig.~\ref{fig_400mu}. The
degree of linear polarization $P_{lin}=\sqrt{P_1^2+P_2^2}$ is
96.8(1.7)\% in average.

It is foreseen in the near future to change to synthetic type IIa diamond plates
for which QWP deviation angles as low as 20 arcsec can be used
\cite{Sca09,Oki01}. Right now, the minimum deviation angles for the
type Ib diamond plates used at P09 are 40 arcsec \cite{Fra13}. 
Type IIa synthetic diamonds produced by the temperature gradient
growth method at high temperature and high pressure are better quality
than type Ib diamonds regarding lattice strain and crystal defect
structures as well as regarding the amount of impurities and
inhomogeneities \cite{Hae05}. A previous study has shown that while synthetic type IIa
diamonds show a typical broadening of the Darwin width of up to 5 arcsec
over the whole crystal, synthetic type Ib diamonds show a width
broadening of up to 20 arcsec \cite{Sum97}.
A direct benefit
will be the larger usable energy range of each crystal and a higher
transmission rate at the lowest deviation angles. The type IIa diamond
phase-plates will cover the energy range from 3.2 to 14
keV. A set of 10 $\mu$m thin silicon phase plates will cover the
energy range between 2.7 and 3.2 keV \cite{Gou96,Bou12}. An additional
double phase-retarder setup allowing operation of two plates in series in
the 90\degree\ geometry has been designed to fit at the optical table in
P09-EH1 in front of the transfocator and the beam monitor for variable
polarization in P09-EH1 and P09-EH2. This device is not in vacuum and
will accommodate the thick type IIa diamond plates ($t > 600 \mu$m).

\subsection{Mirrors}

The mirrors in P09-OH are 1~m long and 130~mm wide fused silica
(SiO$_2$) mirrors with averaged slope errors over 800~mm length of $<
2.5$~$\mu$rad (rms) sagittal, $<0.8$~$\mu$rad (rms) meridional and surface
roughness $< 0.25$~nm (rms). 
In order to preserve parallel beam, the
first mirror deflects vertically up, the second vertically down. The
first mirror (Pilz-Optics, Oberkochen, Germany), has two cylinders
with sagittal radii of curvature of $r_1=88$~mm and a flat section in
between the cylinders.  Half of the flat section and one cylinder are
coated with 40~nm palladium. The second mirror (SESO, Aix-en-Provence,
France) has one cylinder with a sagittal radius of curvature
$r_2=167$~mm and two flat sections at the sides, where one of them is
coated with palladium. In addition, the second mirror is equipped with
a bender allowing meridional radii of curvature down to $R=5$~km for
vertical focusing. By combining different settings of the two mirrors
as shown in Table~\ref{mirror_specs}, it is possible to efficiently
suppress higher harmonics in the energy range from 6 to 24 keV in
P09-EH1 and 7 to 32 keV in P09-EH2 and at the same time to focus the
beam into P09-EH1, P09-EH2 and P09-EH3 by varying the glancing angle.

\begin{table}
\caption{\label{mirror_specs} Specifications of the mirror system for
  different glancing angles (GA). $r_1$ and $r_2$ specify the radii
  for sagittal focusing and $R$ the radius for meridional focusing. Focus
  positions of 76.7~m, 86.5~m and 93~m correspond to sample positions
  in P09-EH1, P09-EH2 and P09-EH3, respectively. The focus position at
  90.1~m corresponds to the position of the intermediate focus at the
  end of P09-EH2 (see section~\ref{lenses}). Numbers in brackets refer
  to cut-offs using the Pd-coated parts.}
\begin{tabular}{lllllll}
Focus pos. & Coating & $r_1$  & $r_2$  & $R$  & GA  & Cutoff\\
(m)  & & (mm) & (mm) & (km) & mrad & (keV)\\\hline
unfocused & SiO$_2$ & flat & flat & flat & variable & variable \\\hline
76.7  & SiO$_2$ (Pd) & 88 & flat & 11.9 & 2.65 & 11.5 (24)\\
 & SiO$_2$ & flat & 167 & 6.0 & 5.28  & 5.7\\\hline
86.5  & SiO$_2$ (Pd) & 88 & flat & 19.4 & 2.13  & 14.3 (31)\\
   & SiO$_2$ & flat & 167 & 9.7 & 4.14 & 7.3 \\\hline
90.1 & SiO$_2$ (Pd)& 88 & flat & 22.8 & 2.00 & 15.1 (32)\\
   & SiO$_2$  & flat & 167 & 11.1 & 3.90 & 7.9 \\\hline
93.0 & SiO$_2$ (Pd)& 88 & flat & 23.1 & 1.95& 15.7 (34)\\
      & SiO$_2$  & flat & 167 & 12.2 & 3.70& 8.3  \\
\end{tabular}
\end{table}

\begin{figure}
\caption{a) Vertical edge scan over the focused beam at the sample
  position in P09-EH1 at 8~keV using the mirrors as the focusing
  device in the 2.65~mrad configuration
  (c.f. Table~\ref{mirror_specs}). The vertical beam size is
  29~\micron. The derivative of the fit to the edge function (dotted
  line) is shown as full line. b) Wire scan (inverted) using a
  5.8~\micron\ wire over the beam, refocused by compound refractive
  lenses, positioned 1.9~m before the sample position in EH1 at
  76.7~m, with a virtual source at 93~m (prefocused by the
  mirror). The width of 4~\micron\ is derived from a fit of a
  Lorentzian convoluted by the Gaussian representing the wire
  thickness. }\label{focus_crl}
%\scalebox{0.55}{\includegraphics{focus.eps}}
\scalebox{0.55}{\includegraphics{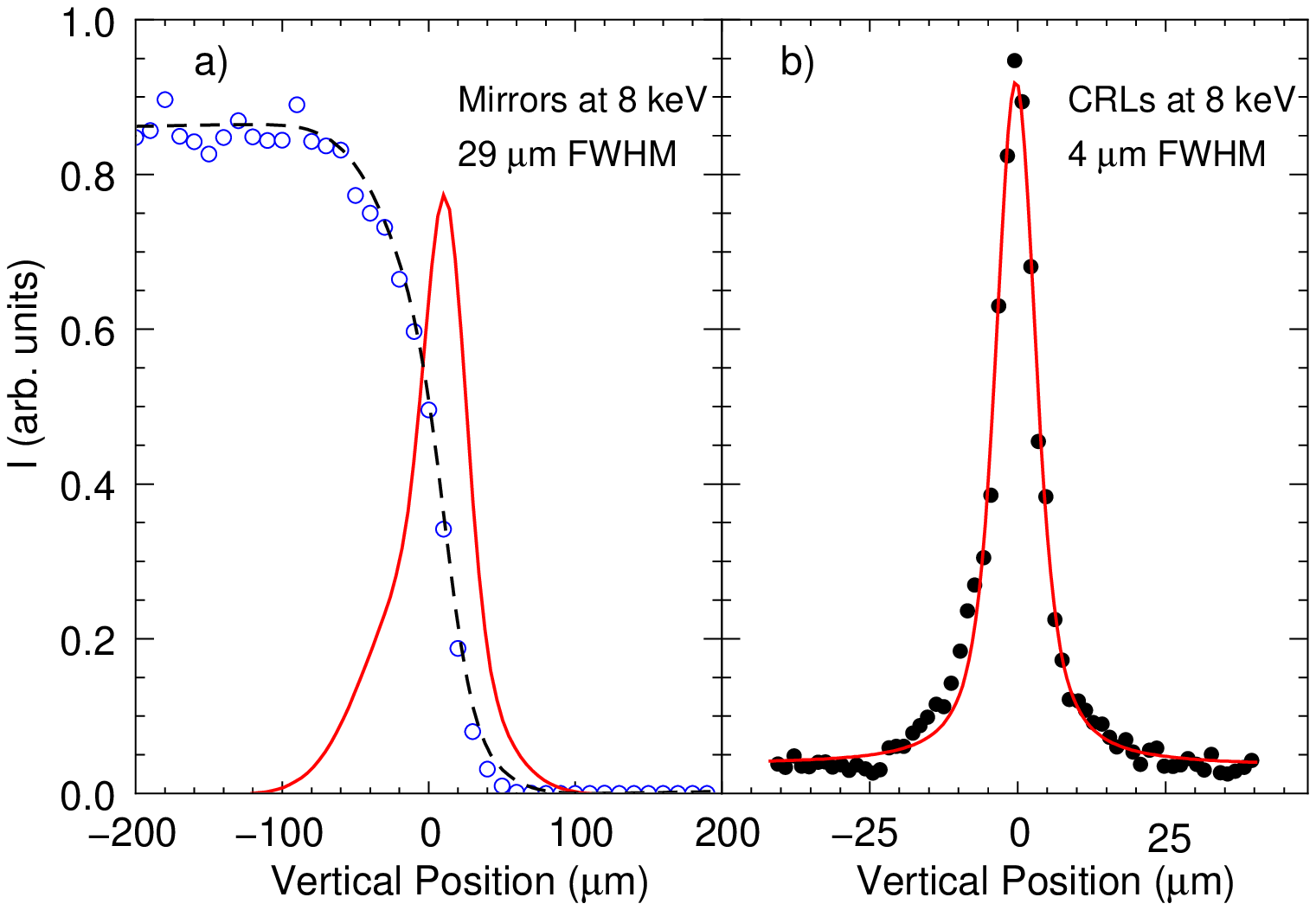}}
\end{figure}

The mirrors are placed in separate UHV chambers. Yaw and pitch
rotations and the translations perpendicular to the beam are
externally applied to the whole chamber. The pneumatic bender is also
external and the bending mechanically transferred into the chamber to
the mirror through bellows. The focus size of the mirror reflected beam at the sample position in
P09-EH1 is $150\times 30$~\micron\ (FWHM).  A vertical knife-edge scan
over the focused beam at the sample position in P09-EH1 is shown in
Fig.~\ref{focus_crl}a. The vertical width corresponds to a local meridional slope error $<0.5$~$\mu$rad (rms) showing the
high quality of some significant mirror regions.

Calculated reflected intensities as function of
energy of the different optical surfaces at different glancing angles
are shown in Fig.~\ref{fig_mirror}. Because of the fixed distance
between the mirrors, the beam height varies with the glancing angle,
which means that the components after the mirrors have to be adjusted
in height accordingly. After both mirrors, removable CVD diamond
screens are located for beam alignment. The intensity of the mirror reflected
beam at the sample position was determined at 8~keV to be $2\cdot
10^{13}$~photons/s using a calibrated photo-diode. This corresponds well to the
calculated value.

The high incident flux is very important when phase plates or CRLs are
used in front of the sample, since these reduce the incident flux
considerably and flux hungry experiments such as non-resonant magnetic scattering
would not be possible otherwise. However, great care has to be
taken for experiments at low temperatures, since beam heating due to
the high photon flux on the sample may affect the sample temperature
considerably. Depending on the type of sample (metal, insulator) and
the temperature range, a reduction of the incident beam intensity
using attenuators in front of the sample might be necessary.

\subsection{Micro focusing with compound refractive lenses}
\label{lenses}

A movable in-vacuum lens changer (transfocator) is positioned in
P09-EH1 or in P09-EH2 at $1.9\pm 0.15$~m in front of the sample
position. The transfocator consists of eight pneumatically actuated
assemblies of CRLs varying in number and radius in order to allow
focusing between 2.7 and 24~keV. It is mounted on motorized pitch and
yaw rotations and x,z translations to align the optical axis of the
CRLs parallel to the beam. It can also be translated by 300~mm along
the beam for fine tuning of the focal point. The beam is pre-focused
to 90.1 m from the source by the mirrors (glancing angle 2.0 or
3.9~mrad) in order to adjust the beam size to the 0.9~mm aperture of
the CRLs. The beam is then refocused to the sample position by the
CRLs resulting in a demagnification factor of about 8:1. Focal sizes
of 4~\micron\ vertical and 50~\micron\ horizontal (FWHM) are achieved,
as is shown in Fig.~\ref{focus_crl}b, which should be compared to
calculated values of 4~\micron\ vertical and
20~\micron\ horizontal. The efficiency for the focusing decreases
with decreasing energy. While at 8~keV, the intensity loss is about a
factor of 3, the overall intensity loss between unfocused and focused
beam is increasing to factors of 20 at 3.5~keV and 25 at 3~keV mainly due to 
small impurities in the Be-lenses.

%%%%%%%%%%%%%%%%%%%%%%%%%%%%%%%%%%%%%%%%%%%%%%%%%%%%%%%%%%%%%%%%%%%%%%%%%%%%
\section{First experimental hutch (EH1) and 6-circle diffractometer}

The first optical element in P09-EH1 is a QBPM with the same
specifications as the one in the optics hutch. A UHV compatible high
precision slit system just after the QBPM is used to define the beam
\cite{Dom07}.  Following the slit system, a moveable CVD diamond screen
is located. A 20~\micron\ diamond window with an opening of 3~mm, which is
mounted in a CF-flange, separates the ring vacuum
from the beamline vacuum. The QPBM, the slit system, the diamond screen
and the diamond window are mounted on a granite table equipped with a
horizontal and vertical translation to align them in the beam. An
in-line electron Time of Flight (eToF) 
polarization monitor using photo-electrons from a gas target
will be positioned on the same table in the near future
\cite{Ilc09}. This device will allow online monitoring of the incident
polarization during polarization manipulation of the incident beam using
the phase retarder setup.
\begin{figure}
%\scalebox{0.33}{\includegraphics{mod4536_comb.ps}}
\scalebox{0.9}{\includegraphics{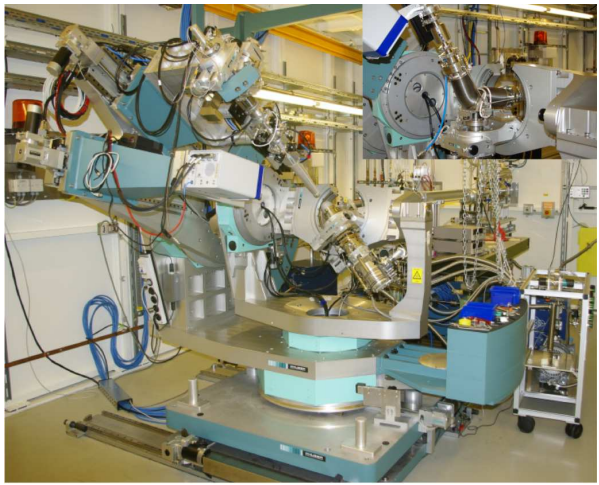}}
\caption{6-circle diffractometer with open $\chi$-circle. On the
  double detector arm, a point detector together with a polarization
  analyzer stage and beam collimating slits as well as a Pilatus 300k pixel
  detector are mounted. Off-setting the scattering angle $\delta$ by
  $25\degree$ allows switching from measurements with the point
  detector to measurements with the 2D detector. The 1.7 K cryocooler
  is mounted on the $\phi$-cradle. It is equipped with Be domes. The
  inset shows the windowless vacuum shroud as replacement for the
  outer beryllium dome for measurements at energies below 4 keV.}
\label{fig_diff_eh1}
\end{figure}

Downstream from the granite table a $0.9\times 1.8$~m$^2$ optical table is
positioned on which the vacuum beam path with different components is
mounted. A mounting system based on Newport X95 rails and carriers and
KF40 flanges allows a very flexible modification of the arrangement of
components. All components are designed to operate under vacuum down
to 10$^{-6}$ mbar. The equipment available on the optical table are
the additional silica mirror and the transfocator. A beam monitor is
positioned after the transfocator. For the beam monitor, a motorized
translation is used to insert different thicknesses of glassy carbon
foils in the beam (20, 60 and 100\micron).  The foils are rotated by
45\degree\ both towards the beam and the ring plane. Two PIPS diodes
with acceptance angles of about 10\degree\ positioned $90\degree$
apart are used to monitor the intensities scattered horizontally and
vertically. The monitor is extensively used to determine the QWP and
HWP positions of the phase plates. The next component is an absorber
box. Using pneumatic actuators, 12 different foils can selectively be
inserted into the beam path, attenuating the beam up to at least 10 orders
of magnitude in the whole energy range of 2.7- 30 keV. The last
component before the diffractometer is an in-vacuum slit system
(JJ-Xray A/S, Kgs. Lyngby, Denmark) followed by a 100~\micron\ thick
Be-window.
\begin{table}
\caption{\label{table_diff_eh1} Parameters of the 6-circle
  diffractometer in P09-EH1. Spheres of confusion (SOC) are obtained
  for loaded sample stage and detector arm. The motor name conventions
  are taken from Ref.~\onlinecite{You99}.}
\begin{tabular}{lcccl}\hline
Motor & Range & Accur.  & SOC & comment \\
 &  (deg) &  (arcsec) & ($\mu m$) &  \\\hline
$\theta$ & $-10$/$+90$ & 0.18 & 30 & vertical $\theta$\\ 
$\delta$ & $-30$/$+180$ & 0.36 & 30 & vertical $2\theta$\\ 
$\mu$ & $\pm 30$ & 0.18 & 25 & horizontal $\theta$\\ 
$\gamma$ & $-20$/$+180$\footnote{dependent on position of $\mu$} & 0.36 & 60 & horizontal $2\theta$ \\ 
$\chi$ & $\pm 90$ & 0.36 & 15 & without cryostat \\ 
& $\pm 48$ & & 30 & with cryostat \\ 
$\phi$ & 0 - 360 & 0.18 & 5 &\\\hline
\end{tabular}
\end{table}

The main instrument in P09-EH1 is a high precision ``4S+2D''
6-circle diffractometer (Huber Diffraktionstechnik) equipped with 4
circles for the sample ($\chi$, $\phi$, $\theta$, $\mu$) and 2 for the
detector ($\delta$, $\gamma$) (Fig.~\ref{fig_diff_eh1}), all circles
being independent. It allows both, vertical and horizontal diffraction. 
The $\chi$-circle has an opening of 15 cm on the
top, allowing for large vertical scattering angles. The
$\phi$ cradle is equipped with the Huber 512.12M motorized xyz-cryostat
carrier permitting a full 360\degree\ rotation. The sample stage can
sustain up to 12 kg load. The reinforced detector arm consists of two
motorized linear translation stages separated vertically by an angle of
25\degree. The point detector setup with beam collimation and analyzer
and the 2D detector are both mounted simultaneously, as shown in
Fig.~\ref{fig_diff_eh1}. The diffractometer is standing on a
xz-table with a $+60/-30$ mm vertical and $\pm 20$~mm horizontal
translation and a pitch of $\pm 1.5$\degree. The angular ranges along
with spheres of confusion for the different circles are shown in
Table~\ref{table_diff_eh1}.

The following closed-cycle cryostats are available at P09: a 4 to
450~K ARS DE-202SG cryocooler (8~K base temperature on sample), a 6 to
800~K ARS CS-202AG cryocooler, a 1.7 to 300~K ARS DE-302 cryocooler
(3~K base temperature at sample in the absence of exchange gas) and a vibration free
4~K Cooltran He-flow cryostat. The cryostats are limiting the $\chi$
angular range as shown in Table~\ref{table_diff_eh1}.  They are
equipped with Be-domes: an outer vacuum dome with wall thickness of
0.5 mm (D1) and a second dome acting as heat shield with wall
thickness of 0.38 mm (D2). A third Be dome for exchange gas with a
wall thickness of 0.38 mm (D3) is available for all
cryostats. For low energy experiments, the external dome can be
replaced by a windowless vacuum shroud to reduce absorption, which
becomes relevant for energies below 4~keV (see inset of
Fig.~\ref{fig_diff_eh1}). At 2.8~keV, the intensity gain, when
omitting the outer dome, amounts already to a factor of 10 for both
the incident and diffracted beam as shown in
Table~\ref{table_absorption_be}. The windowless vacuum shroud has an
angular range of two times 50 degrees. The range in $\chi$ is $\pm
5$\degree. An azimuthal rotation is possible through the double O-ring
coupling at the cryostat.  

\begingroup
\begin{table}
\caption{\label{table_absorption_be} Calculated X-ray transmission for
  different configurations of Be-domes (Displex cryostats) and
  Be-windows (14~T magnet) as function of energy. The first two rows
  show the transmission without and with the outer dome (D1) in
  addition to the the radiation shield (D2) for the Displex
  cryostats. Rows 3 and 4 show the transmission when the dome for
  exchange gas (D3) is mounted in addition. Rows 5 and 6 show the
  transmission for the 14~T magnet without and with
  $^3$He-insert. Actual values for the low energies might be increased
  considerably due to impurities.}
\begin{tabular}{lccccc}\hline
Be (mm) & 8 keV & 6 keV & 4 keV & 2.8 keV &  Configuration \\\hline
0.76 & & 0.71  & 0.32 & $3.3\cdot 10^{-2}$ &  D2 \\
1.76 & 0.71 & 0.46 & 0.07 & $3.9\cdot 10^{-4}$ &  D1, D2 \\\hline
1.52 & & 0.51 & 0.11 &  $1.1\cdot 10^{-3}$ &  D2, D3 \\
2.52 & 0.61 & 0.33 & 0.02 & $1.3\cdot 10^{-5}$ &  D1, D2, D3\\\hline
4.0 & 0.45 & 0.18 & $3\cdot 10^{-3}$ & $1.7\cdot 10^{-8}$ & Magnet\\
6.0 & 0.31 & 0.07 & $1\cdot 10^{-4}$ & $2.3\cdot 10^{-12}$& M., $^3$He-ins.\\\hline
\end{tabular}
\end{table}
\endgroup

On the detector arm, a Huber tube cross slit system 3002.60.M followed
by a Huber cross slit system 3002.70.M are mounted to define the beam
path between sample and detector. The connection to the polarization analyzer stage
is done using a double O-ring coupling in order to allow an
in-vacuum rotation of the complete analyzer stage around the scattered
beam axis ($\eta'$-rotation) and a complete in-vacuum beam path
starting at the Kapton window behind the sample up to the
detector. The range of $\eta'$ is 0 to -150 degrees. The analyzer crystal is
mounted with the analyzer Bragg angle of $\theta_{pol}=45 \pm 5 \degree$ towards the 
diffracted beam. The beam diffracted by the analyzer is recorded
by the detector positioned at a
scattering angle $2\theta_{pol} = 90\pm 8\degree$. The analyzer
crystals are attached to the goniometer using kinematic mounts allowing for a fast
change of the crystals. The height and roll of the analyzer crystals can be
adjusted using Attocube piezo stages. A selection of the analyzer
crystals available at the beamline together with most relevant
absorption edges is shown in Table~\ref{ana_crystals}. A Huber
2-circle goniometer 415 can be mounted on the 2-theta arm instead of
the polarization analyzer stage for experiments requiring higher
resolution.

Cyberstar NaI-scintillation detectors and APD-detectors with ESRF-type
amplifiers are available as point detectors at P09 (FMB-Oxford). An
energy-dispersive silicon-drift VORTEX EX90 detector (SII
NanoTechnology USA, Northridge, CA, USA) with 25~\micron\ Be window is
used as a fluorescence detector. A Pilatus 300k 2D pixel detector
(Dectris, Baden, Switzerland) is also
available at the beamline.

\begin{table}
\caption{\label{ana_crystals} Crystals available for polarization
  analysis from 2.7 up to 13 keV. Photon energies for $2\theta =
  90^{\circ}$ diffraction angle are given for each crystal and
  reflection, together with some relevant absorption edges close by.}
\begin{tabular}{lcccl}\hline
Crystal  & (h k l) & d (\AA) & E (90\degree) (keV)& Absorption edges  \\\hline
Graphite & (0 0 2) & 3.358 & 2.61 & Mo(\ltwo),Ru(\lthree)\\
TbMnO$_3$ & (0 2 0) & 2.928 & 2.99 & Ru(\ltwo) \\
Au & (1 1 1) & 2.338 & 3.75 & U(\mfour, \mfive)\\
Cu & (1 1 1) & 2.084 & 4.21 & \\ 
Graphite & (0 0 4) & 1.679 & 5.22 & V(K) \\
Mo & (2 0 0) & 1.574 & 5.57 & Ce(\lthree)\\
Al & (2 2 0) & 1.432 & 6.12 & Cr(K), Pr(\lthree), Nd(\lthree)\\
Cu & (2 2 0) & 1.276 & 6.84 & Mn(K), Fe(K), Eu(\lthree), \\
 & & & &Nd(\ltwo), Sm(\lthree)\\
Au & (2 2 2) & 1.169 & 7.50 & Tb(\lthree), Sm(\ltwo)\\
Graphite & (0 0 6) & 1.119 & 7.83 & Dy(\lthree), Gd(\ltwo)\\
%Au & (4 0 0) & 1.020 & 8.60 & Dy (\ltwo)\\
Cu & (2 2 2) & 1.042 & 8.41 & Dy(\ltwo), Tm (\lthree)\\
Pt & (4 0 0) & 0.981 & 8.94 & Cu(K), Ho(\ltwo), Yb(\lthree)\\
Graphite & (0 0 8) & 0.839 & 10.44 & Os(\lthree)\\
Au & (3 3 3) & 0.779 & 11.25 & Ir(\lthree) \\ 
Graphite & (0 0 10) & 0.671 & 13.05 & Ir(\ltwo), Os(\ltwo)\\\hline
\end{tabular}
\end{table}

%%%%%%%%%%%%%%%%%%%%%%%%%%%%%%%%%%%%%%%%%%%%%%%%%%%%%%%%%%%%%%%%%%%%%%%%%%

\section{Second experimental hutch (EH2) and heavy-load 6-circle diffractometer}

\begin{figure}
%\scalebox{0.6}{\includegraphics{magnet_eh2.eps}}
\scalebox{1.7}{\includegraphics{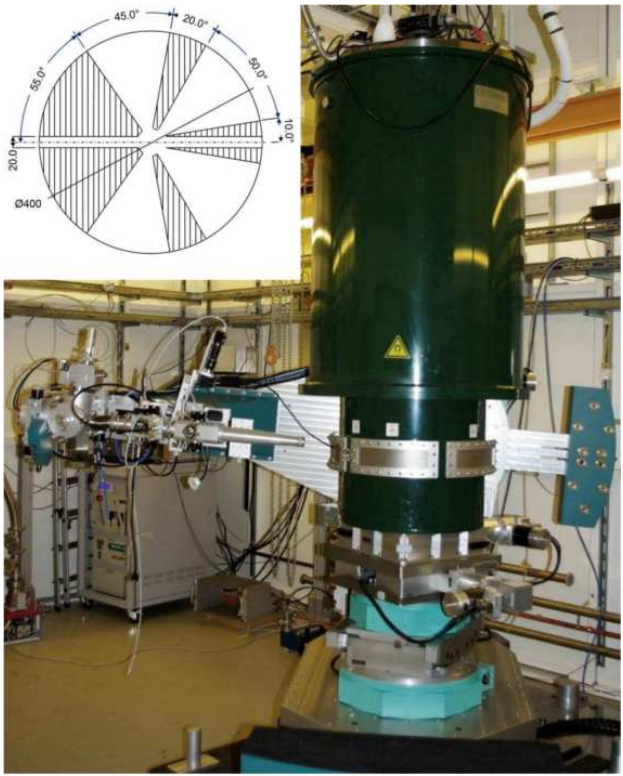}}
\caption{Heavy load 6-circle diffractometer with 14~T magnet and
  polarization analyzer. The insert shows the window arrangement of
  the 14~T cryomagnet in the horizontal plane.}
\label{diff_eh2}
\end{figure}

P09-EH2 is equipped with a non-magnetic heavy load 6-circle
diffractometer in horizontal Psi geometry (Huber Diffraktions\-technik)
which is capable of carrying a load of up to 650 kg. There are two
independent circles for the detector arm: $\gamma$ (vertical scattering
angle) and $\delta$ (horizontal scattering angle), and three for the
sample: $\omega$ (horizontal $\theta$), $\chi$ and $\phi$
(Fig.~\ref{diff_eh2}). An additional degree of freedom is $\mu$, the
limited vertical pitch of $\pm 1 \degree$ of the whole diffractometer
that moves both sample and detector. The $\chi$-cradle has limited
movement of $\pm 7 \degree$ for loads less than 200~kg while only $\pm
3 \degree$ for larger weights. The xy- and z-stages allow a $\pm
5$~mm and $\pm 10$~mm horizontal and vertical translation,
respectively. The angular ranges along with the spheres of confusion
for the different circles are shown in Table~\ref{table_diff_eh2}. The
detector arm carries a polarization analyzer stage and two slit
systems for beam collimation identical to the setup in P09-EH1. The
slit systems can be removed, making room for the Pilatus 300k area
detector.  In front of the diffractometer, a multi purpose non-magnetic
optical table is located (ADC Inc., Lansing, USA), on which the vacuum
beam path is mounted. Also here, the transfocator from P09-EH1 can be
inserted.

A vertical field 14 T split-pair superconducting magnet (Cryogenic
Ltd., London, UK) is available in P09-EH2. It is equipped with a
variable temperature insert (VTI) providing temperatures in the range
of 1.8 to 300~K in normal operation. A $^3$He insert is also available
and allows access to sample temperatures as low as 300 mK. The maximum power
deposited on the sample at 8~keV is 26~mW which has to be compared to
the cooling power of the $^3$He-stage of $<1$~mW. This implies that
measurements at these temperatures for a reasonable time are possible
only with reduced beam intensities by more than 2 orders of
magnitude. 

The X-ray windows around the magnet, at the VTI and the
$^3$He insert are from 1~mm thick beryllium each, in order to reduce
absorption of the X-ray beam to a minimum. Nevertheless, below 6~keV
intensity is reduced already significantly (see
Table~\ref{table_absorption_be}). This limits the usable energy range
to energies above 5~keV. The vertical opening of the X-ray windows is
$\pm 5$\degree\ around an aperture of 10 mm. Combining the $\phi$ and
$\theta$ angles and the rotation of the sample probe within the
cryostat, the horizontal access has no dark angles and large
horizontal scattering angles of $2\theta$ up to 170\degree\ can be
reached. The horizontal window distribution is shown in the inset of
Fig.~\ref{diff_eh2}.

\begin{table}
\caption{\label{table_diff_eh2} Parameters of the horizontal
  Psi-diffractometer in P09-EH2. Spheres of confusion are
  obtained for sample stage loaded with 150~kg and loaded detector arm. The
  motor name conventions are taken from
  Ref.~\onlinecite{You99}. Compared to the 6-circle diffractometer in
  P09-EH1, vertical and horizontal rotations are swapped.}
\begin{tabular}{lcccl}\hline
Motor & Range & Accur. & SOC  & comment \\
 & (deg) &  (arcsec) &  ($\mu m$) &  \\\hline
$\theta$ & $\pm 180$ & 0.18 & 10 & horizontal $\theta$\\ 
$\delta$ & $-20$/$+180$ & 0.36 & 15 & horizontal $2\theta$\\ 
$\mu$ & $\pm 1$ & 0.18 & - & vertical $\theta$\\ 
$\gamma$ & $-30$/$+180$ & 0.36 & 70 & vertical $2\theta$ \\ 
$\chi$ & $\pm 7$ &0.36 & 15 & $<200$~kg \\ 
& $\pm 3$ & &  & max. 650~kg load\\ 
$\phi$ & 0 - 360 & 0.18 & $<5$ & \\\hline
\end{tabular}
\end{table}

%%%%%%%%%%%%%%%%%%%%%%%%%%%%%%%%%%%%%%%%%%%%%%%%%%%%%%%%%%%%%%%%%%%%%%%%%
\section{Experiment control}

All relevant devices at the beamline are controlled by TANGO,
including a diffractometer server for moves in reciprocal space.
SPECTRA/ONLINE is the data acquisition and beamline control software used at
P09 as a command line interface between the user and the Tango devices
\cite{Kra07,Alf11}. A variety of moves and scans are available.

Some Tango devices special to P09 are mentioned in the following. The
{\sl multiple motors} device allows the definition of slave devices to the
energy. For the slave devices, a relation between energy and an
angular or linear movement is defined, which then moves the slave
device whenever the energy is changed. As slave devices, the {\sl
  undulator}, the {\sl phase retarder} and the {\sl polarization
  analyzer} can be selected. Like this, energy scans over a peak with arbitrary 
polarization (e.g. circular) is feasible. The {\sl mirror} TANGO device allows
selection between the several different mirror configurations and moves
the two mirrors into position.  The {\sl phase retarder} TANGO
devices are used to select between the different phase plates
available on each tower and to move the phase plates to 1/2, 1/4 or 1/8
wave plate condition.  The {\sl diffractometer} device relates the
reciprocal space defined by the crystal structure of the sample
mounted on the diffractometer to the angular movements of the
diffractometer according to the Busing-Levi convention
\cite{Bus67,You99}. The HKL library was developed by F. Picca at SOLEIL. 
It allows selection between different diffractometer geometries and
several different modes of diffraction (bisecting, constant-phi,
constant-psi, z-axis) \cite{Pic11}.

%%%%%%%%%%%%%%%%%%%%%%%%%%%%%%%%%%%%%%%%%%%%%%%%%%%%%%%%%%%%%%%%%%%%%%%%%%%%

%%%%%%%%%%%%%%%%%%%%%%%%%%%%%%%%%%%%%%%%%%%%%%%%%%%%%%%%%%%%%%%%%%%%%%%%%%%%

\section{Conclusion and outlook}

Beamline P09 has been operational since mid 2010 for external
users. Diffraction experiments in both experimental hutches have been
performed and show that the equipment is performing according to
specification. The investigation of weak non-resonant magnetic
reflections in SmFeAsO shows that very weak signals can be measured
reliably \cite{Nan11}. Full polarization analysis, performed at the Ho
\lthree-edge of HoFe$_3$(BO$_3$)$_4$ could be used to show the spiral
magnetic order along the c-axis \cite{Shu12}. From azimuthal scans at
the Ir \lthree-edge of Sr$_2$IrO$_4$, the ordering of the Ir moments in
the basal plane could be determined and be related to the ordering in
Ba$_2$IrO$_4$ \cite{Bos13}. Application of electric fields and
reciprocal space mapping was used for the investigation of the
electric field induced superlattice reflections from the
polar layer at the LaAlO$_3$/SrTiO$_3$ interface \cite{Roe13}.

Above mentioned representative experiments indicate that P09 combines
all the possibilities required for scattering experiments and can be
used to perform state of the art experiments in this field. Concerning
photon flux, polarization control, polarization analysis, focusing
capabilities and high magnetic fields it compares favorably with other
experimental stations at $3^{rd}$-generation synchrotron sources like
APS (4ID-D, 6-ID-B) and Spring-8 (BL22XU) as well as Diamond (I16),
ESRF (XMAS) and BESSY (MagS).

The type IIa diamond phase-plates, together with the installation of the
eToF polarization monitor in the near future will provide means to vary and control
polarization in a wide energy range and observe any
irregularities during polarization scans, respectively.

\begin{acknowledgements}{The authors would like to acknowledge the staff of the experiment
  control group FS-EC, M. Hesse and the beamline technology group
  FS-BT, M. Tischer and the undulator group FS-US and J. Spengler and
  the general infrastructure group FS-TI. We also want to acknowledge discussion and advice
  during the planning of the beamline from O.H. Seeck, J.C. Lang,
  C. Detlefs, P. Thompson and L. Bouchenoire. Also, we would like to
  thank K. Pflaum, D. Samberg, A. Gade and R. D\"oring for engineering
  and technical support.}
\end{acknowledgements}

     % References are at the end of the document, between \begin{references}
     % and \end{references} tags. Each reference is in a \reference entry.

%\bibliography{/home/strempfr/tex/bib/p09_instrument,/home/strempfr/tex/bib/books,/home/strempfr/tex/bib/magscatt,/home/strempfr/tex/bib/anom_scatt,/home/strempfr/tex/bib/phaseret,/home/strempfr/tex/bib/tbmno3,/home/strempfr/tex/bib/strempfer,/home/strempfr/tex/bib/sonstig}

     %-------------------------------------------------------------------------
     % TABLES AND FIGURES SHOULD BE INSERTED AFTER THE MAIN BODY OF THE TEXT
     %-------------------------------------------------------------------------

     % Postscript figures can be included with multiple figure blocks

\end{document}